# Calculation of distributed system imbalance in condition of multifractal load

Kirichenko Lydmila
Dept. of Applied Mathematics
Kharkiv National University of Radio Electronics
Kharkiv, Ukraine
Lyudmyla.kirichenko@nure.ua

Radivilova Tamara
Dept. of Telecommunication Systems
Kharkiv National University of Radio Electronics
Kharkiv, Ukraine
tamara.radivilova@nure.ua

Ivanisenko Igor
Dept. of Applied Mathematics
Kharkiv National University of Radio Electronics
Kharkiv, Ukraine
Ihor.ivanisenko@nure.ua

*Abstract*—**The method of calculating a distributed system imbalance based on the calculation of node system load was proposed in the work. Calculation of node system load is carried out by calculating the average coefficient of utilization of CPU, memory, bandwidth of each system node, taking into account multifractal properties of input data flows. The simulation of the proposed method for different multifractal parameters was conducted. The simulation showed that the characteristics of multifractal traffic have a considerable effect on the system imbalance. Using the proposed method allows to distribute requests across the servers so that the deviation of the load servers from the average value was minimal, which allows to get higher system performance metrics and faster processing flows.**

## I. INTRODUCTION

The investigation of processes in information networks have shown that the network traffic has scale invariance property (self-similarity) [1-3]. Self-similar traffic has a special structure that continuing on many scales - in the realization are always a certain number of very large bursts at relatively low average traffic level. The presence of the self-similarity property at information flows transferred customer has a great influence on the performance of distributed systems [4-9]. A particularly important role it plays for services providing the transmission of multimedia and real-time traffic [6, 8]. One of the methods of providing quality of service is a load balancing through optimal allocation of tasks between system nodes [6- 14]. Thus, the actual problem is the calculation of load nodes and distributed system imbalance in self-similar input flow [15-20].

The proposed load balancing method is based on the complex internal and external monitoring methods. Using an external monitoring system allows to periodically test the network to determine the most congested segments. Using the internal monitoring of node status allows to get an objective information about load of node and assembly components.

Criteria for load node is proposed CPU utalization, memory, and bandwidth for flows with ifferent classes of service. To take account of the potential procesing power of each node of the resource, it is proposed to use weights. This method takes into account the multifractal properties of the incoming information flow. Using an integrated approach for balancing the load would lead to increased performance distributed computing system.

Each server's load calculation is carried out by calculating the average CPU, memory, bandwidth utilization of each server, and then distributed system load calculated [10]. The obtained information about loading used for the balancing process to determine the occurrence of imbalance and to determine a new tasks distribution by calculating the scope of work is needed for moving tasks [15, 17, 20]. The criteria for evaluating distributed system load are below .

## II. EVALUATION CRITERIA OF DISTRIBUTED SYSTEM LOAD

1. Average CPU utilization $CPU_i^u$ of a single server $i$ defined as the averaged CPU utilization during an observed period $T$.

Similarly, the average utilization of $r$ memory $RAM_i^r(T)$, network bandwidth of $k$ channel $Net_i^k(T)$ of server $i$ can be defined.

2. Average utilization of all CPUs in a system. Let $CPU_i^{n_i}$ be the total number of CPUs of server $i$,

$$CPU_u^{All} = \sum_i^N CPU_i^u CPU_i^{n_i} / \sum_i^N CPU_i^{n_i} \qquad (1)$$

where N is the total number of servers in a system, $n_i$ is amount of CPU of server $i$. Similarly, the average utilization of memory $RAM_i^{m_i}$, network bandwidth $Net_i^{k_i}$ of server $i$, all memories $RAM_r^{All}$, and all network bandwidth $Net_k^{All}$ in a system can be defined.

3. The imbalance value of all CPUs, memories, and network bandwidth. Using dispersion formulas, the imbalance value of all CPUs in distributed system is defined as

$$ISL_{CPU} = \sum_i^N (CPU_i^u - CPU_u^{All})^2 \qquad (2)$$

Similarly, imbalance values of memory $ISL_{RAM}$ and network bandwidth $ISL_{Net}$ can be calculated. Then total imbalance values of all servers in a system is given by

$$IBL_{tot} = \sum_i^N IBL_i \qquad (3)$$

4. Let introduce the complex value of load imbalance $SIL_i$ of server $i$, which takes into account all three server resource. Using the formula for calculating dispersion as a measure of irregularity, the integrated value of load imbalance of server $i$ can be defined as:

$$SIL_i = a(CPU_i^u - CPU_u^{All})^2 + b(RAM_i^r - RAM_r^{All})^2 + \\ + c(Net_i^k - Net_k^{All})^2 \qquad (4)$$

Parameters $a, b, c$ represent weighting coefficients for the processor, memory and network bandwidth, respectively, which are selected by experimentation, so that $a+b+c=1$ and depend on the tasks and the system structure. $SIL_i$ is used to indicate load imbalance level by comparing the coefficient of CPU utilization, memory, and bandwidth.

5. Then, the total imbalance value of all servers in the system is defined as:

$$ISL_{tot} = \sum_i^N SIL_i / N \qquad (5)$$

6. Efficiency is defined as the average load on any server.

Thus, the method of complex measurements of general level of integrated system imbalance has been developed for scheduling resources, as well as the average level of each server imbalance.

For analysis of the proposed method of calculation of the distributed system imbalance is necessary to conduct simulations. To the entrance of a distributed system the flow of tasks, which has a self-similar and multifractal properties must be supplied. Generation model of multifractal stochastic process are presented in [5].

### III. CHARACTERISTICS OF SELF-SIMILAR AND MULTIFRACTAL PROCESSES

Stochastic process $X(t)$, $t \geq 0$, with continuous real-time variable is said to be self-similar of index $H$, $0 < H < 1$, if for any value $a > 0$ processes $X(at)$, and $a^{-H}X(at)$, have same finite-dimensional distributions:

$$Law\{X(at)\} = Law\{a^H X(t)\}. \qquad (6)$$

The notation $Law\{\cdot\}$ means finite distribution laws of the random process. Index $H$ is called Hurst exponent. It is a measure of self-similarity or a measure of long-range dependence of process. For values $0,5 < H < 1$ time series demonstrates persistent behaviour. In other words, if the time series increases (decreases) in a prior period of time, then this trend will be continued for the same time in future. The value $H = 0,5$ indicates the independence (the absence of any memory about the past) time series values. The interval $0 < H < 0,5$ corresponds to antipersistent time series: if a system demonstrates growth in a prior period of time, then it is likely to fall in the next period.

The moments of the self-similar random process can be expressed as $E[|X(t)^q|] = C(q) \cdot t^{qH}$ where the quantity $C(q) = E[|X(1)|^q]$.

In contrast to the self-similar processes (6) multifractal processes have more complex scaling behavior $Law\{X(at)\} = Law\{M(a) \cdot X(t)\}$ where $M(a)$ is random function that independent of $X(t)$. In case of self-similar process $M(a) = a^H$.

For multifractal processes the following relation holds $E[|X(t)|^q] = c(q) \cdot t^{qh(q)}$ where $c(q)$ is some deterministic function, $h(q)$ is generalized Hurst exponent, which is generally non-linear function. Value $h(q)$ at $q = 2$ is the same degree of self-similarity H. Generalized Hurst exponent of monofractal process does not depend on the parameter q: h(q)=H.

As a characteristic of heterogeneity multifractal data flow in the work was proposed to calculate range of generalized Hurst exponent $\Delta h = h(q_{min}) - h(q_{max})$. For monofractal processes generalized Hurst exponent is independent of parameter q, and is a straight line: h(q)=H, Δh=0. The greater heterogeneity of the process, ie. large number of bursts present in the traffic, the greater the range Δh.

### IV. USING THE TEMPLATE

Software which allows the simulation of work load balancing systems for different parameters of the input multifractality data flow and by using the proposed method of calculation of the distributed system imbalance was developed. In order to form an input load using the model described in [5], multifractal traffic with variable multifractality parameters was generated.

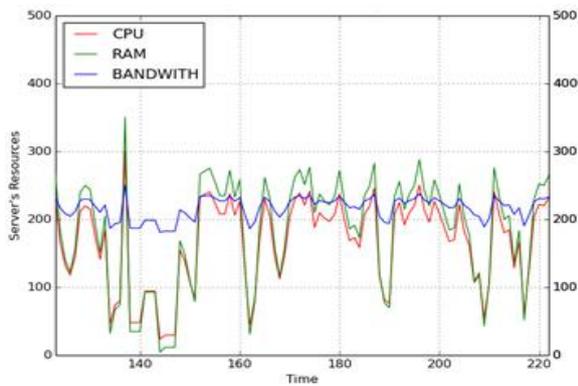

Fig. 1. Schedule server resources load H=0.6, Δh=1.5

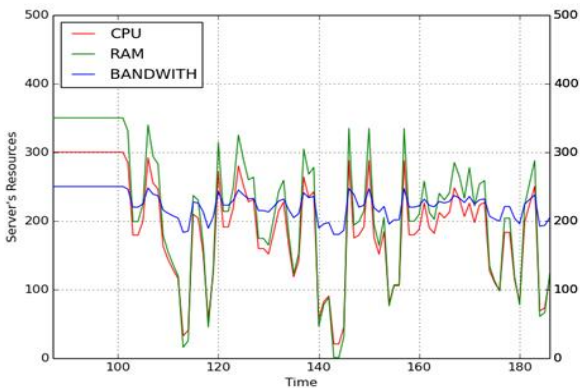

Fig. 2. Schedule server resources load H=0.6, Δh=2.5

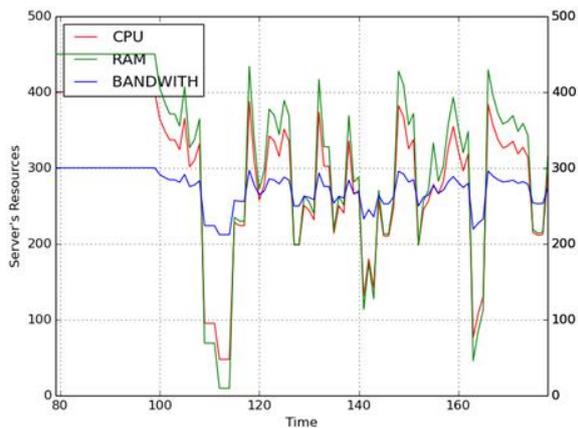

Fig. 3. Schedule server resources load H=0.9, Δh=2.5

Fig. 1-3 shows the change of processors, memory, and channel load. Parameters a,b,c (4) that indicate the weights for the processor, memory, and bandwidth, were selected equivalent.

In the first case (Fig. 1) the load balancer receives generated traffic with parameter H=0.6 and range of the generalized Hurst exponent Δh=1.5. Fig. 2 shows a load system for traffic with stronger long dependence (H=0.9) and the same inhomogeneity in the first case. Fig. 2 shows the results of the modeling work balancer for large values of Hurst parameter H=0.9, and the range of the generalized Hurst exponent Δh=2.5.

Researches shows that the system imbalance depends essentially on the multifractal traffic characteristics. For small H values and small inhomogeneities the load balancing system reaches equilibrium and imbalance value to tends to zero. By increasing Hurst parameter over time the imbalance of the system does not decay and load balancing system doesn't comes to equilibrium. For large values of Hurst parameter and large heterogeneity load balancing system is in an unstable state and imbalance value is changed several times, resulting in a maximum resources loading.

## Conclusion

Load balancing method based on estimating the load of distributed system nodes proposed in this work. The average CPU, memory, and bandwidth load are calculated when evaluating the nodes load, based on measured load by accounting system or the operating system monitor. This allows to calculate the system imbalance, the average duration of work and the efficiency of using system resources.

The simulation results showed that characteristics of multifractal traffic significantly affect to system imbalance. For small values of Hurst exponent and a small traffic heterogeneity imbalance value tends to zero and load balancing system reaches equilibrium. For large values of the Hurst exponent and the heterogeneity the load balancing system is always in an unstable state, which leads to a maximum load of resources. Using the proposed method for load balancing, taking into account information about server and system state, allows to load balancer to allocate server, which is able to best deal with the processing of multifractal task flow.

## References

[1] L. Kirichenko, T. Radivilova, Analysis of network performance under selfsimilar system loading by computer simulation, Bionics intelligence, №1, 2008, pp.158-160.

[2] O. I. Sheluchin, S. M. Smolskiy, A. V. Osin, Self-Similar Processes in Telecommunications, New York : John Wiley & Sons, 2007, pp. 320.

[3] J.W. Kantelhardt, Fractal and multifractal time series, Mathematics of complexity and dynamical systems, 2012, pp. 463-487.

[4] L. Kirichenko, I. Ivanisenko, T. Radivilova, Investigation of Self-similar Properties of Additive Data Traffic, CSIT-2015 X-th International Scientific and Technical Conference «Computer science and information technologies», Lviv, UKRAINE, 14 – 17 September, 2015, pp. 169-172.

[5] L. Kirichenko, T. Radivilova, E. Kayali, Modeling telecommunications traffic using the stochastic multifractal cascade process, Problems of Computer Intellectualization ed. K. Markov, V. Velychko, O. Voloshin, Kiev–Sofia: ITHEA, 2012, pp. 55–63.

[6] V. Cardellini, E. Casalicchio, M. Colajanni, A performance study of distributed architectures for the quality of web services, Proceedings of the 34th Conference on System Sciences, Vol. 10. 2001, pp. 213-217.

[7] V. Cardellini, M. Colajanni, P. S. Yu, Dynamic Load Balancing on Web-server Systems, IEEE Internet Computing, vol.3(3), 1999, pp. 28-39.

[8] A.A. Dort-Goltz, Development and research of a method of balancing the traffic in packet communication networks. Thesis for Ph.D, St. Petersburg State University of Telecommunications, 2014, P. 168.

[9] S. Keshav, An Engineering Approach to Computer Networking, Addison-Wesley, Reading, MA, 1997, pp. 215-217.

[10] Wenhong Tian, Yong Zhao, Optimized Cloud Resource Management and Scheduling: Theories and Practices, Morgan Kaufman, 2014, P. 284.

[11] Xing-Guo Luo, Jing Liu, Xing-Ming Zhang, Fan Zhang, Bai-Nan Li, Job Scheduling Model for Cloud Computing Based on Multi-Objective Genetic Algorithm, IJCSI International Journal of Computer Science, v.10(1), № 3, 2013, pp. 134-139.

[12] Hisao Kameda, Lie Li, Chonggun Kim, Yongbing Zhang, Optimal Load Balancing in Distributed Computer Systems, Springer, Verlag London Limited, 1997, P. 238.

[13] Kalyani Ghuge, Minaxi Doorwar, A Survey of Various Load Balancing Techniques and Enhanced Load Balancing Approach in Cloud Computing, International Journal of Emerging Technology and Advanced Engineering, Volume 4, Issue 10, 2014, pp. 410-414.

[14] M. Mendonca, B.A.A. Nunes, X.-N. Nguyen, K.Obraczka, T. Turletti, A Survey of software-defined networking: past, present, and future of programmable networks, Communications Surveys & Tutorials, IEEE, Vol.16(3), 2013, pp. 1617-1634.

[15] Rudra Koteswaramma, Client-Side Load Balancing and Resource Monitoring in Cloud, International Journal of Engineering Research and Applications (IJERA), Vol.2(6), 2012, pp. 167-171.

[16] Dhinesh Babu L.D., P. Venkata Krishna, Honey bee behavior inspired load balancing of tasks in cloud computing environments, Applied Soft Computing, Volume 13, Issue 5, 2013, pp. 2292–2303,.

[17] E.I. Ignatenko, V.I. Bessarab, I.V. Degtyarenko, An adaptive algorithm for monitoring network traffic cluster in the load balancer, Naukovi pratsi DonNTU, Vol.21(183), 2011, pp. 95-102.

[18] Zhihao Shang, Wenbo Chen, Qiang Ma, Bin Wu, Design and implementation of server cluster dynamic load balancing based on OpenFlow, Awareness Science and Technology and Ubi-Media Computing (iCAST-UMEDIA), 2013, pp. 691 – 697.

[19] Martin Randles, David Lamb, A. Taleb-Bendiab, A Comparative Study into Distributed Load Balancing Algorithms for Cloud Computing, IEEE 24th International Conference on Advanced Information Networking and Applications Workshops, 2010, pp. 551-556.

[20] Thomas Erl, Robert Cope, Amin Naserpour, Cloud Computing Design Patterns, Prentice Hall, Ed.1st., 2015, p.592.
.